\documentclass[preprint]{revtex4-1}

\usepackage{graphicx}
\usepackage{amssymb}
\usepackage{amstext}
\usepackage{amsbsy}
\usepackage{subfigure}
\usepackage{amsmath}

\begin{document}

\title{Generalization of the Schwarz-Christoffel mapping to multiply connected polygonal domains}

\author{Giovani L. Vasconcelos$^{1,2}$ }\address{$^{1}$Department of Mathematics, Imperial College London, 180 Queen's Gate, London SW7 2AZ, United Kingdom\\
$^{2}$Departamento de F\'{\i}sica,  Universidade Federal de Pernambuco, 50670-901, Recife, Brazil}



\begin{abstract}
A generalization of the  Schwarz-Christoffel mapping to multiply connected polygonal domains is obtained by making a combined use of two preimage domains, namely, a rectilinear slit domain and a bounded circular domain. The  conformal mapping from the circular domain to the polygonal region is  written as an indefinite integral whose integrand consists of a product of powers of the Schottky-Klein prime functions, which is the same irrespective of the preimage slit domain, and a prefactor function that depends on the choice of the rectilinear slit domain. A detailed derivation of the mapping formula is given for the case where the preimage slit domain is the upper half-plane with radial slits. Representation formulae for other canonical slit domains are also obtained but they are more cumbersome in that the prefactor function contains arbitrary parameters in the interior of the circular domain.
\end{abstract}

\maketitle

\section{Introduction}

\label{sec:1}

The Schwarz-Christoffel mapping 
to polygonal domains is an important result in the theory of complex-valued functions and one that  finds numerous applications in applied mathematics, physics, and engineering 
\cite{Driscoll}.  The applicability of Schwarz-Christoffel formula is  nonetheless limited by the fact that it pertains only to simply connected polygonal domains. Therefore, there has long been considerable interest in extending the Schwarz-Christoffel transformation to multiply connected polygonal domains.
 
 Recently, explicit formulae for generalized Schwarz-Christoffel mappings from a circular domain to both bounded and unbounded multiply connected polygonal regions have been obtained using two equivalent approaches. Using reflection arguments, DeLillo and collaborators \cite{DeLillo2004,DeLillo2006} derived infinite product formulae for the generalized Schwarz-Christoffel mappings, whereas  Crowdy \cite{BoundedMCSC, UnboundedMCSC} used a function-theoretical approach to obtain a mapping formula  in terms of the Schottky-Klein prime function associated with the circular domain. (The equivalence of the two methods was established in \cite{DeLillo2006}.) It is important to point out that both these formulations entail an implicit choice of a multiply connected rectilinear slit domain  in an auxiliary preimage  plane, whereby each rectilinear boundary segment of this preimage domain is mapped to a polygonal boundary (see below). 
The introduction of a preimage domain with rectilinear boundaries makes it possible to write down an explicit expression for the derivative of the desired conformal mapping, which can then be integrated to yield a generalised Schwarz-Christoffel formula.
It is thus clear that different choices of rectilinear slit domains in the auxiliary plane will result in different formulae for the generalised Schwarz-Christoffel mapping.
Therefore, a derivation of alternative representation formulae is of some interest. 

In this paper a general formalism to construct 
conformal mappings of multiply connected polygonal domains is presented. The method applies to both bounded and unbounded polygonal domains 
but emphasis will be given to the bounded case as formulated in \S\ref{sec:2}. 
A key ingredient in the approach described herein is the
introduction of 
a multiply connected rectilinear slit domain in an auxiliary $\lambda$-plane, 
such that each rectilinear boundary in the $\lambda$-plane is mapped to a
polygonal boundary in the $z$-plane. It is  advantageous to have rectilinear boundaries in the preimage domain  because the derivative of the respective conformal mapping has constant argument on each of domain's boundaries.  
To derive an expression for the mapping derivative, it is expedient to 
 introduce a slit map from a bounded multiply connected circular domain in an auxiliary $\zeta$-plane to the rectilinear slit domain in the $\lambda$-plane. Using the properties of the Schottky-Klein prime function  associated with the circular domain, as summarized in \S\ref{sec:3} and \S\ref{sec:4},
 it is then possible to obtain an explicit expression for the derivative of the mapping from the circular domain to the polygonal region. 
The desired conformal mapping is then written in final form as an indefinite integral  whose  integrand consists of  i) a product of powers of the Schottky-Klein prime functions
and  ii) a  prefactor function,   $S(\zeta)$, that depends on the choice of the rectilinear slit domain in the $\lambda$-plane. 

A detailed derivation of the prefactor $S(\zeta)$ is  given in \S\ref{sec:5} for the case where the  rectilinear slit domain consists of the upper half-plane cut along radial segments.
This preimage domain turns out to be the most convenient one not only in that it naturally extends to multiply connected domains the usual treatment of simply connected polygonal regions but most importantly because it yields a simpler formula for the generalized Schwarz-Christoffel mapping. 
As further illustration of the method presented here, 
the corresponding formulae for two other canonical slit domains (including the case considered by Crowdy \cite{BoundedMCSC}) are  derived in \S\ref{sec:6}. These alternative representations have, however, the inconvenience of containing certain arbitrary parameters in the interior of the circular domain. For completeness, the case of unbounded polygonal domains is briefly discussed in \S\ref{sec:7}, after which our main results and conclusions are summarized in \S\ref{sec:8}.

\section{Mathematical formulation}

\label{sec:2}

\subsection{Bounded polygonal regions}
\label{sec:2a}

\begin{figure}
\begin{center}
\subfigure[$\lambda$-plane \label{fig:1a}]{\includegraphics[width=0.35\textwidth]{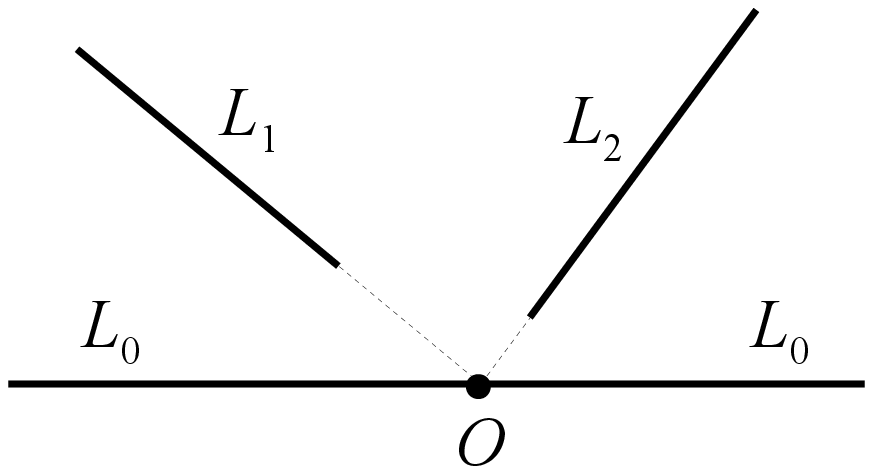}}
\qquad\qquad
\vspace{0.4cm}
\subfigure[$z$-plane \label{fig:1b}]{\includegraphics[width=0.25\textwidth]{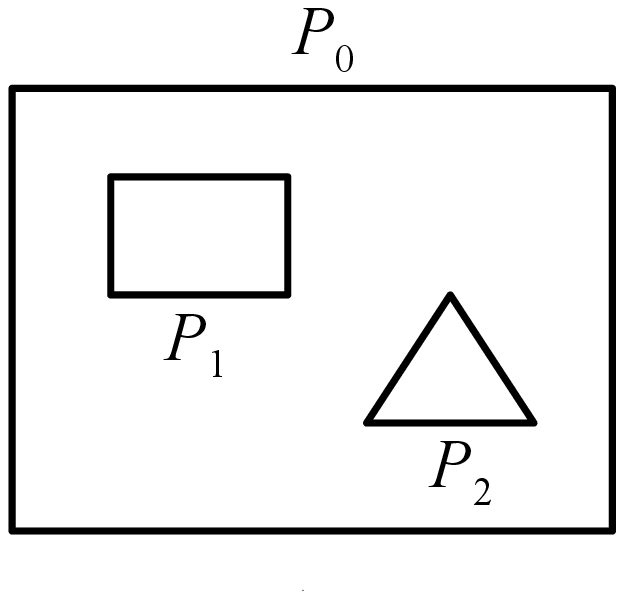}}\\
\subfigure[$\zeta$-plane\label{fig:1c}]{\includegraphics[width=0.4\textwidth]{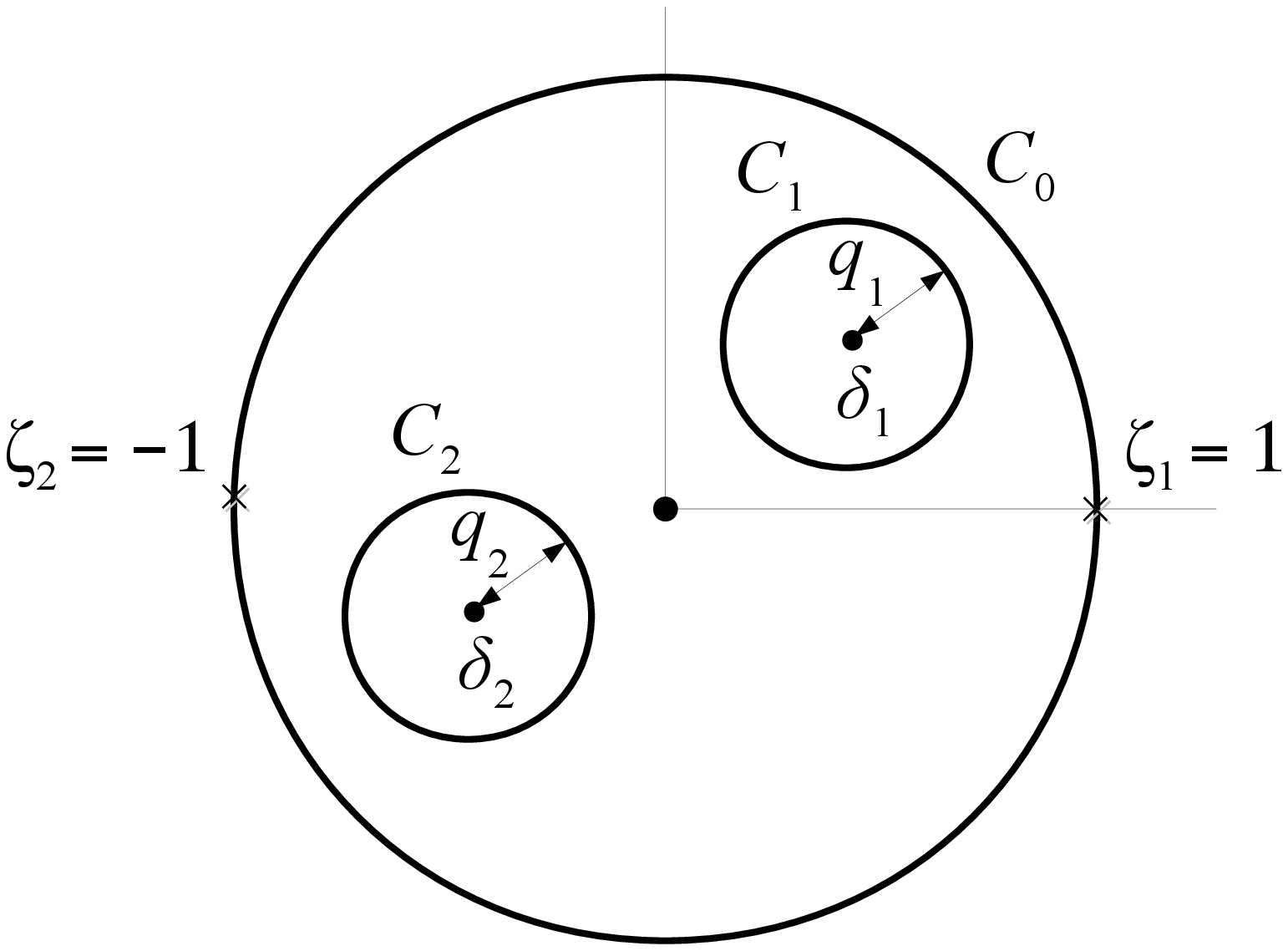}}
\caption{ (a) Radial slit domain in the upper half-$\lambda$-plane;   (b) bounded polygonal domain in the $z$-plane;  and (c) circular domain in the   $\zeta$-plane.}
\end{center}
\end{figure}

Let us consider a target region $D_z$ in the $z$-plane that is  a bounded $(M+1)$-connected polygonal domain. The outer boundary polygon is denoted by $P_0$ and  the $M$  inner polygons by $P_j$, $j=1,...,M$; see figure \ref{fig:1b}. Let the vertices at polygon $P_j$, $j=0, 1,...,M$, be denoted by $z_k^{(j)}$,  $k=1,...,n_j$, and let $\pi \alpha_{k}^{(j)}\in[0,2\pi]$  be the interior angles at the respective vertices. Following the notation of Driscoll \& Trefethen \cite{Driscoll}, we write
\begin{align}
\pi\alpha_k^{(j)}=\pi(\beta_k^{(j)}+1), \qquad k=1,...,n_j,
\end{align}
where  $\pi \beta_k^{(j)}$ represents the turning angle at  vertex $z_k^{(j)}$, so  that the parameters  $\beta_k^{(j)}$  must satisfy the following relations:
\begin{align}
\sum_{k=1}^{n_0}\beta_k^{(0)}=-2, \qquad \sum_{k=1}^{n_j}\beta_k^{(j)}=2, \quad j=1,...,M.
\label{eq:bj}
\end{align}

\subsection{Radial slit domains in the upper half-plane}
 
 \label{sec:2b} 
Let us now consider  a domain $D_\lambda$  in a subsidiary $\lambda$-plane consisting of the upper half-plane with $M$  slits pointing towards the origin.  Denote by $L_0$  the real axis in the $\lambda$-plane  and by $L_j$, $j=1,...,M$, the $M$ radial slits; see figure \ref{fig:1a}. 
We seek a conformal mapping  $z(\lambda)$  from the radial slit domain $D_\lambda$  in the upper half-$\lambda$-plane  to the polygonal domain $D_z$ in the $z$-plane,  where the real $\lambda$-axis is mapped to the outer polygon $P_0$ and each radial slit $L_j$  is mapped to an inner polygon $P_j$. 

If we denote by $\lambda_{k}^{(j)}$  the preimages in the $\lambda$-plane of the vertices $z_k^{(j)}$ on polygon $P_j$, $j=0,1,...,M$, then  $z(\lambda)$ must have a branch point at  $\lambda=\lambda_{k}^{(j)}$ such that
\begin{equation}
\frac{dz}{d\lambda}\approx \textrm{constant}\cdot (\lambda-\lambda_{k}^{(j)})^{\beta_k^{(j)}} \quad\mbox{for}\quad \lambda\to \lambda_{k}^{(j)}.
\label{eq:zb}
\end{equation}
Furthermore, 
the derivative $dz/d\lambda$ must have  constant arguments on each boundary segment in the $\lambda$-plane, that is,
\begin{equation}
\arg\left[\frac{dz}{d\lambda}\right]= \textrm{const.} \quad\mbox{for}\quad \lambda\in L_j, \ j=0,1,...,M.
\label{eq:argdzb}
\end{equation}
This follows from the fact  that both $\arg[d\lambda]$ and $\arg[dz]$ are constant on the respective boundaries in the $\lambda$- and $z$-planes. In the case of simply connected polygonal regions, conditions (\ref{eq:zb}) and (\ref{eq:argdzb}) can be easily satisfied by writing $dz/d\lambda$ as a product of monomials of the type $(\lambda-\lambda_{k}^{(j)})^{\beta_k^{(j)}}$, 
leading to the  well-known Schwarz-Christoffel formula \cite{Fokas}.
But for multiply connected domains  condition (\ref{eq:argdzb})
represents a major obstacle in deriving an explicit formula  for  $dz/d\lambda$, for no longer is it obvious how to construct a function that has constant argument on the multiple rectilinear boundaries of the domain $D_\lambda$.
This obstacle can nonetheless be overcome by 
introducing a conformal mapping, $\lambda(\zeta)$, from a circular domain $D_\zeta$ in an auxiliary $\zeta$-plane 
to the slit domain $D_\lambda$. This allows us to write an explicit expression for the derivative $dz/d\lambda$ in terms of the $\zeta$  variable, as  will be seen in \S\ref{sec:5}.

\subsection{Bounded circular domains}

\label{sec:2c}

Let $D_\zeta$ be a circular domain in the $\zeta$-plane consisting of the unit  disc with $M$ smaller nonoverlapping disks excised from it. Label the unit circle by $C_0$ and  the $M$ inner circular boundaries by $C_1,...,C_M$, and denote the centre and radius of the circle $C_j$ by $\delta_j$ and $q_j$, respectively. For convenience, we introduce the notation $\delta_0=0$ and $q_0=1$ for the unit circle. A schematic of $D_\zeta$ is shown in figure \ref{fig:1c}. 

Now let  $\lambda(\zeta)$ be a conformal mapping from the circular domain $D_\zeta$ to the domain $D_\lambda$ defined above, 
where the unit circle, is mapped to the real axis in the $\lambda$-plane and the interior circles $C_j$ map to the slits  $L_j$, $j=1,...,M$. Furthermore, let  the point $\zeta=1$ map to the origin in the $\lambda$-plane and the point $\zeta=-1$ map to infinity; see figures \ref{fig:1a} and \ref{fig:1c}. In an abuse of notation, we shall  write  $z(\zeta)\equiv z(\lambda(\zeta))$ 
for the associated conformal mapping  from $D_\zeta$ to the polygonal region $D_z$, where  the unit circle  $C_0$ maps to the outer polygonal boundary $P_0$ and the interior circles $C_j$ map to the inner polygons $P_j$; see figures \ref{fig:1b} and \ref{fig:1c}.
If we denote by  $a_{k}^{(j)}$  the preimages in the $\zeta$-plane of the vertices $z_k^{(j)}$ on polygon $P_j$, then condition (\ref{eq:zb}) can be recast as
\begin{equation}
{z_\lambda}(\zeta)\approx \textrm{constant}\cdot (\zeta-a_{k}^{(j)})^{\beta_k^{(j)}} \quad\mbox{for}\quad \zeta\to a_{k}^{(j)},
\label{eq:za}
\end{equation}
where ${z_\lambda}(\zeta)$ denotes the derivative  $dz/d\lambda$ expressed in terms of the  $\zeta$-variable. Similarly, the requirement (\ref{eq:argdzb}) becomes
\begin{equation}
\arg\left[{z_\lambda}(\zeta)\right]= \textrm{const.} \quad\mbox{for}\quad \zeta\in C_j, \ j=0,1,...,M.
\label{eq:argdz}
\end{equation}
As first noticed by Crowdy  \cite{BoundedMCSC}, albeit in a somewhat different and less general formulation,  it is possible to write  an explicit expression for $z_\lambda(\zeta)$ that satisfies conditions (\ref{eq:za}) and (\ref{eq:argdz}) by exploiting the  properties of the Schottky-Klein prime function associated with the circular domain $D_\zeta$. 
Once  an expression for $z_\lambda(\zeta)$ is obtained,
 a corresponding expression for $z_\zeta(\zeta)$ follows from the 
chain rule which can then be integrated yielding the desired  conformal mapping $z(\zeta)$, as will be shown in \S\ref{sec:5}. Prior to this, however, a brief overview of the Schottky-Klein prime function is  in order. 

\section{Schottky  groups and the Schottky-Klein prime function}

\label{sec:3}

Consider the bounded circular domain $D_\zeta$ defined above; see figure \ref{fig:1c}. Introduce the following M\"obius maps:
\begin{equation}
\theta_j(\zeta)=\delta_j+\frac{q_j^2 \zeta}{1-\bar{\delta}_j\zeta}, \quad j=0,1,...M,
\label{eq:mobius}
\end{equation}
where a bar denotes complex conjugation. For $\zeta \in C_j$, it is easy to establish the following  relations:
 \begin{align}
 {\zeta}={\theta_j}(1/\bar\zeta)  \qquad \Longleftrightarrow\qquad  \bar{\zeta}=\bar{\theta}_j(1/\zeta),
 \label{eq:zbar}
 \end{align}
 where in the second identity we introduced the conjugate function $\bar\theta(\zeta)=\overline{\theta(\bar\zeta)}$ .

Now let $C'_j$,  $j=1,...,M$,  denote the reflection of the circle $C_j$ in the unit circle $C_0$. One can  readily verify that every $\theta_j(\zeta)$ maps the interior of the circle $C_j'$ onto the exterior of the circle $C_j$. 
Thus, the set  $\Theta$ consisting of all compositions of the maps $\theta_j(\zeta)$, $j=1,...,M$, and their inverses defines a  classical Schottky group \cite{Baker}. The  region in the complex plane exterior to all of the circles $C_j$ and $C_j'$ is called the {\it fundamental region}, $F$, of the Scotkky group $\Theta$. Given a Scotkky group $\Theta$, we can associate a Schottky-Klein prime function, denoted by $\omega(\zeta,\alpha)$, for any two distinct points $\zeta$ and $\alpha$ in the fundamental region $F$.
 
The Schottky-Klein prime function has the following infinite product representation \cite{Baker}:
\begin{equation}
\omega(\zeta,\alpha)=(\zeta-\alpha)\prod_{\theta \in \Theta''} \frac{(\zeta-\theta(\alpha))(\alpha-\theta(\zeta))}{(\zeta-\theta(\zeta))(\alpha-\theta(\alpha))},
\label{infiprod}
\end{equation}
where the subset $\Theta'' \subset \Theta$ consists of all compositions of the maps $\theta_j(\zeta)$ and $\theta^{-1}_j(\zeta)$, excluding the identity and taking only an element or its inverse (but not both). For example, if $\theta_1(\theta^{-1}_2(\zeta))$ is included in the set $\Theta''$, then $\theta_2(\theta^{-1}_1(\zeta))$ must be excluded.

Let us now recall some useful  functional identities involving the Schottky-Klein prime function.  Firstly,  note that by definition the Schottky-Klein prime function is antisymmetric in its two arguments:
\begin{equation}
\omega(\zeta,\gamma)=-\omega(\gamma,\zeta).
\label{refid0}
\end{equation}
Secondly, for the Schottky-Klein prime function associated with the circular domain $D_\zeta$ the following functional relation
\begin{equation}
\bar{\omega}\left(1/\zeta,1/\gamma \right)=-\frac{1}{\zeta \gamma}\omega(\zeta,\gamma)
\label{id1}
\end{equation}
holds \cite{KRtheory}. 
A third important relation of the Schottky-Klein prime function is as follows. 
Let $\zeta_1$, $\zeta_2$, $\gamma_1$, and $\gamma_2$ be four  points in $F$, then we have 
\begin{equation}
\frac{\omega(\theta_j(\zeta_1),\gamma_1)}{\omega(\theta_j(\zeta_2),\gamma_2)}=\frac{\beta_j(\gamma_1,\gamma_2)}{\beta_j(\zeta_1,\zeta_2)}
\left(\frac{1-\bar\delta_j\zeta_2}{1-\bar\delta_j\zeta_1}\right)
\frac{\omega(\zeta_1,\gamma_1)}{\omega(\zeta_2,\gamma_2)},
\label{idrat}
\end{equation}
where
\begin{equation}
\beta_j(\zeta,\gamma)=\exp \left[2 \pi \mathrm{i}  (v_j(\zeta)- v_j(\gamma))\right],  \quad j=1,...,M.
\label{eq:beta}
\end{equation}
Equation (\ref{idrat}) follows  from a related expression given in ch.~12 of Baker \cite{Baker} for the ratio ${\omega(\theta_j(\zeta),\gamma)}/{\omega(\zeta,\gamma)}$;  see also related discussion in the monograph by Hejhal \cite{Hejhal}.
 Here the functions $\{v_j(\zeta)~|~j=1,...,M\}$
are the $M$ integrals of the first kind of the Riemann surface associated with the domain $D_\zeta$. These are analytic but not single-valued functions everywhere in $F$. (They can be made single-valued by inserting cuts connecting each pair of circles $C_j$ and $C_j'$; see Baker \cite{Baker}.) 
Each function $v_j(\zeta)$
has constant imaginary part on the circles $C_1,...,C_M$, and zero imaginary part on the unit circle $C_0$, that is,
 \begin{align}
 {\rm Im}[v_j(\zeta)]=Q_{jl} \qquad\mbox{for}\qquad \zeta \in C_l, \ l=0,1,...,M, 
 \label{eq:nuj}
\end{align}
where $Q_{jl}$ is a real constant, with $Q_{j0}=0$ \cite{CM2006}.

As particular cases of  (\ref{idrat}) we have
\begin{equation}
\frac{\omega(\theta_j(\zeta),\gamma_1)}{\omega(\theta_j(\zeta),\gamma_2)}=\beta_j(\gamma_1,\gamma_2)\frac{\omega(\zeta,\gamma_1)}{\omega(\zeta,\gamma_2)},
\label{idrat2}
\end{equation}
\begin{equation}
\frac{\omega(\theta_j(\zeta_1),\gamma)}{\omega(\theta_j(\zeta_2),\gamma)}=
\frac{1}{\beta_j(\zeta_1,\zeta_2)}
\left(\frac{1-\bar\delta_j\zeta_2}{1-\bar\delta_j\zeta_1}\right)
\frac{\omega(\zeta_1,\gamma)}{\omega(\zeta_2,\gamma)}.
\label{idrat3}
\end{equation}
We also note for later use that for $\zeta_1, \zeta_2\in C_l$, $l=0,1,...,M$, the function
$\beta_j(\zeta,\gamma)$ satisfies the following relations
\begin{align}
|\beta_j(\zeta_1,\zeta_2)|&=1, \label{eq:b1}
\\
\beta_j(1/\bar\zeta_1,1/\bar\zeta_2)&={\beta_j(\zeta_1,\zeta_2)}.
\label{eq:barbeta}
\end{align}
Identity (\ref{eq:b1})
follows immediately from (\ref{eq:beta}) and (\ref{eq:nuj}). To derive  (\ref{eq:barbeta}), first notice that from  (\ref{eq:nuj}) we have
\begin{align}
v_j(\zeta)=\overline{ v}_j(1/ \zeta),
\label{eq:vjC0}
\end{align}
for $ \zeta\in C_0$ and everywhere else in $F$ by analytic continuation. On the other hand, for $\zeta\in C_l$, $l=1,...,M$,   relation (\ref{eq:nuj}) implies that
\begin{align}
v_j(\zeta)- v_j(1/\bar \zeta)&=2 \mathrm{i} Q_{jl},
\end{align}
where we have used (\ref{eq:vjC0}). This, together with (\ref{eq:beta}), implies (\ref{eq:barbeta}).
Note furthermore that  relations (\ref{idrat2}) and (\ref{idrat3}) trivially hold  for $j=0$, if we define 
\begin{equation}
\beta_0(\gamma_1,\gamma_2)\equiv1.
\end{equation}

\section{Radial slit maps}
\label{sec:4}

In this section, two general classes of functions are defined as ratios of Schottky-Klein prime functions (or of products thereof) in such a way that they have  constant arguments on the circles $C_j$, $j=0,1,...,M$. Because of this property, which will be extensively used in \S\ref{sec:5}, these functions represent radial slit maps defined on the circular domain $D_\zeta$.
 Here there are two cases to consider depending on whether the image radial slit domain is bounded or unbounded.
 
 \subsection{Bounded radial slit domains}

\label{sec:4a}

First define  the  functions
\begin{equation}
F_j(\zeta;\zeta_1,\zeta_2)=\frac{\omega(\zeta,\zeta_1)}{\omega(\zeta,\zeta_2)},  \qquad \zeta_1,\zeta_2\in C_j, \quad j=0,1,...,M.
\label{eq:Fj}
\end{equation}
(These functions were  introduced by Crowdy \cite{BoundedMCSC} as two separate classes of functions; here we adopt a somewhat different notation and give a unified treatment of them.)  An important property of the functions above
is that they have constant argument on all circles $C_l$, $l=0,1,...,M$. 
To see this,  note that for $\zeta\in C_l$ one has 
\begin{flalign*}
F_j(\zeta;\zeta_1,\zeta_2)&=
\frac{{\omega({\theta_l}(1/\bar\zeta),\zeta_1)}}{{\omega({\theta_l}(1/\bar\zeta),\zeta_2)}}
&& \mbox{[from (\ref{eq:zbar})]}\cr 
&={\beta_l (\zeta_1,\zeta_2)}
\frac{\omega(1/\bar\zeta, \zeta_1)}{\omega(1/\bar\zeta, \zeta_2)}
&& \mbox{[from (\ref{idrat2})]}\cr 
&={\beta_l (\zeta_1,\zeta_2)}
\frac{\omega({\theta_j}(1/\bar\zeta_1),1/\bar\zeta)}{\omega({\theta_j}(1/\bar\zeta_2),1/\bar\zeta)}&& \mbox{[from (\ref{refid0}) and (\ref{eq:zbar})]}\cr 
&=\frac{\beta_l (\zeta_1,\zeta_2) }{{\beta_j(1/\bar\zeta_1,1/\bar\zeta_2)}}
\left(\frac{1-\bar\delta_j/\bar\zeta_2}{1-\bar\delta_j/\bar\zeta_1}\right)
\frac{\omega(1/\bar\zeta,1/\bar\zeta_1)}{\omega(1/\bar\zeta,1/\bar\zeta_2)}&& \mbox{[from (\ref{idrat3})]}\cr
&=\frac{\beta_l (\zeta_1,\zeta_2) }{{\beta_j(\zeta_1,\zeta_2)}}
\left(\frac{1-\bar\delta_j/\bar\zeta_2}{1-\bar\delta_j/\bar\zeta_1}\right)
\frac{\omega(1/\bar\zeta,1/\bar\zeta_1)}{\omega(1/\bar\zeta,1/\bar\zeta_2)}.&& \mbox{[from (\ref{eq:barbeta})]}
\end{flalign*}
Taking complex conjugate and using (\ref{id1}) yields
\begin{flalign}
\overline{F_j(\zeta;\zeta_1,\zeta_2)}
&=\frac{\overline{\beta_l (\zeta_1,\zeta_2)}}{\overline{\beta_j(\zeta_1,\zeta_2)}}
\left(\frac{\zeta_2-\delta_j}{\zeta_1-\delta_j}\right)
F_j(\zeta;\zeta_1,\zeta_2),
\label{eq:Fjbar}
\end{flalign}
which implies 
in view of  (\ref{eq:b1}) 
 that 
\begin{align}
\arg \left[F_j(\zeta;\zeta_1,\zeta_2)\right]
= \mbox{const.} \quad\mbox{for}\quad   \zeta\in C_l,  \ l=0,1,...,M.
\label{eq:argF}
\end{align}

Since $F_j(\zeta;\zeta_1,\zeta_2)$ has constant argument on $C_l$, $l=0,1,...,M$, a simple zero at $\zeta=\zeta_1\in C_j$, and a simple pole at $\zeta=\zeta_2\in C_j$, it immediately follows
that this function  maps the circular domain $D_\zeta$ onto a half-plane punctured along $M$ radial segments, where the circle $C_j$ is mapped  to the axis containing the origin whose direction has argument $\arg[F_j(\zeta;\zeta_1,\zeta_2)]$ and the other circles $C_l$, $l\ne j$, are mapped to the slits.  An alternative formula for this mapping in terms of an infinite product was obtained  by DeLillo and Kropf \cite{DeLillo2010}.

From the preceding discussion it is then clear that the conformal mapping defined by
\begin{align}
\lambda(\zeta)&= -\mathrm{i}\, F_0(\zeta;1,-1)\cr
&= -\mathrm{i}\frac{\omega(\zeta,1)}{\omega(\zeta,-1)}
\label{eq:tau0}
\end{align}
maps the circular domain $D_\zeta$ onto the upper half-$\lambda$-plane with $M$ radial slits excised from it, where the points $\zeta=1$ and $\zeta=-1$ are respectively mapped to the origin and infinity in the $\lambda$-plane, the unit circle maps  to the real axis,  and the inner circles map to the radial slits; see figures \ref{fig:1a} and \ref{fig:1c}. 
For later use, we  quote here the derivative of  mapping (\ref{eq:tau0}):
\begin{align}
\frac{d\lambda}{d\zeta} =-\mathrm{i}\frac{ \omega_\zeta(\zeta,1)\omega(\zeta, -1)-\omega_\zeta(\zeta,-1)\omega(\zeta, 1)}{ \omega(\zeta,-1)^2},
\label{eq:lz}
\end{align}
where $\omega_\zeta(\zeta,\alpha)$ denotes the derivative of $\omega(\zeta,\alpha)$ with respect to its first argument.

\subsection{Unbounded radial slit domains}

\label{sec:4b}

Now  consider a second  class of functions  defined by the following ratio:
\begin{equation}
Q(\zeta;\alpha,\beta)=\frac{\omega(\zeta,\alpha)\omega(\zeta,\bar\alpha^{-1})}{\omega(\zeta,\beta)\omega(\zeta,\bar\beta^{\,-1})},
\label{eq:Qj}
\end{equation}
where $\alpha$ and $\beta$ are two arbitrary points in $D_\zeta$.
 Using  arguments analogous to those employed in the \S\ref{sec:4}\ref{sec:4a}, it is easy to verify \cite{BoundedMCSC,CM2006} that 
\begin{align}
\arg \left[Q(\zeta;\alpha,\beta)\right]
= \mbox{const.} \quad\mbox{for}\quad   \zeta\in C_j,  \ j=0,1,...,M.
\label{eq:argQ}
\end{align} 
It then follows that  the  mapping
\begin{align}
\lambda=Q(\zeta;\alpha,\beta)
\label{eq:lQ}
\end{align}
conformally maps $D_\zeta$ to the entire $\lambda$-plane cut along $M+1$  radial slits, where the point $\zeta=\alpha$  is  mapped to the origin in the $\lambda$-plane, the point  $\zeta=\beta$ is mapped to  infinity, and each circle $C_j$, $j=0,1,...,M$, is mapped to  a radial slit in the $\lambda$-plane.

The functions $F_j(\zeta;\zeta_1,\zeta_2)$ and $Q(\zeta;\alpha,\beta)$ defined above 
play an important role  in constructing  conformal mappings to multiply connected polygonal domains, as will become evident in the next section.

\section{Conformal mappings to bounded polygonal domains}

\label{sec:5}

In this section, we  construct  an explicit formula for the conformal mappings, $z(\zeta)$, from the bounded circular domain $D_\zeta$ to a bounded multiply connected  polygonal domain $D_z$, using as an auxiliary tool  the slit map $\lambda(\zeta)$ from $D_\zeta$ to the upper half-$\lambda$-plane with $M$ radial slits.
As  explained in \S\ref{sec:2}\ref{sec:2c}, we first need to obtain an expression for the derivative $z_\lambda(\zeta)$ such that it has: i) the appropriate branch point
at the prevertices $a_k^{(j)}$ 
 and ii) constant argument on the circles $C_j$.

To this end, let $\{\gamma_{1}^{(j)},\gamma_{2}^{(j)} \in C_j | j=0,1,...,M\}$ be a set of arbitrary points on the circles $C_j$.
Using (\ref{eq:bj}), (\ref{eq:Fj}) and (\ref{eq:argF}), it is not difficult to show
that the functions
\begin{align*}
\omega(\zeta,\gamma_{1}^{(0)}) \omega(\zeta,\gamma_{2}^{(0)}) \prod_{k=1}^{n_0} \left[\omega(\zeta,a_{k}^{(0)})\right]^{\beta_k^{(0)}}
\end{align*}
and
\begin{align*}
\frac{\prod_{k=1}^{n_j} \left[\omega(\zeta,a_{k}^{(j)})\right]^{\beta_k^{(j)}}}{\omega(\zeta,\gamma_{1}^{(j)}) \omega(\zeta,\gamma_{2}^{(j)})}, \qquad j=1,...,M,
\end{align*}
all have constant arguments on the circles $C_l$, $l=0,1,...,M$.
Let us then write 
\begin{equation}
\frac{d z}{d\lambda}=\mathcal{B}R(\zeta)\left\{
\frac{ \omega(\zeta,\gamma_1^{(0)})\omega(\zeta, \gamma_2^{(0)})}{\prod_{j=1}^{M} \omega(\zeta,\gamma_{1}^{(j)}) \omega(\zeta,\gamma_{2}^{(j)})}
\prod_{j=0}^{M} {\prod_{k=1}^{n_j} \left[\omega(\zeta,a_{k}^{(j)})\right]^{\beta_k^{(j)}}}\right\}.
\label{eq:SC4}
\end{equation}
where $\mathcal{B}$ is a complex constant and
 $R(\zeta)$ is a correction function to be determined later. Because $\omega(\zeta,\gamma)$ has a simple zero in $\zeta=\gamma$, it  is clear  that $z_\lambda(\zeta)$ 
 has the right branch point at $\zeta=a_k^{(j)}$; see (\ref{eq:za}). Furthermore, it  follows from the preceding discussion that
  (\ref{eq:SC4}) will have constant argument on  the circles $C_j$, so long as  the  function $R(\zeta)$ does so.
  This requirement, together with  additional constraints on $z_\lambda(\zeta)$ concerning the location of its zeros and poles, dictates the specific form of  $R(\zeta)$, as shown next.
  
First note that the points $\{\gamma_{1}^{(j)},\gamma_{2}^{(j)}\in C_j|j=1,...,M\}$  appearing in (\ref{eq:SC4})  must correspond to the preimages in the $\zeta$-plane of the end points of the respective slits in the $\lambda$-plane. This follows from the fact that $d\lambda/d\zeta$ vanishes at the preimages of the slit end points, whereas $dz/d\lambda$ does not; hence,  $z_\lambda(\zeta)$ must have simple poles at these points.
More specifically, the points $\gamma_{1}^{(j)}$ and $\gamma_{2}^{(j)}$, for $j=1,...,M$, are obtained by computing the roots (on the circle $C_j$) of the following equation:
\begin{align}
 \omega_\zeta(\zeta,1)\omega(\zeta, -1)-\omega_\zeta(\zeta,-1)\omega(\zeta, 1)=0,
 \label{eq:gj}
\end{align}
which yields the zeros of  $d\lambda/d\zeta$; see (\ref{eq:lz}).
Note, furthermore, that since $\gamma_{1}^{(0)}$ and $\gamma_{2}^{(0)}$ are arbitrary points on the unit circle at which $z_\lambda(\zeta)$ is nonzero, the terms  containing these points in the numerator of (\ref{eq:SC4}) must  be cancelled out  by an appropriate choice of the function $R(\zeta)$. In addition, $R(\zeta)$  must also  produce a  double zero for  $z_\lambda(\zeta)$ at $\zeta=-1$, since  $d\lambda/d\zeta$  has a double pole at this point; see (\ref{eq:lz}).

These  requirements  can be satisfied by choosing  $R(\zeta)$ of the form
 \begin{align}
 R(\zeta) &= \frac{\omega(\zeta,-1)^2}{\omega(\zeta,\gamma_1^{(0)})\omega(\zeta,\gamma_2^{(0)})},
 \label{eq:R1}
 \end{align}
 which clearly has constant argument on the circles $C_j$; see  (\ref{eq:argF}).
Inserting (\ref{eq:R1}) into (\ref{eq:SC4}) yields
 \begin{equation}
\frac{d z}{d\lambda}=\mathcal{B}
\frac{ \omega(\zeta,-1)^2}{\prod_{j=1}^{M} \omega(\zeta,\gamma_{1}^{(j)}) \omega(\zeta,\gamma_{2}^{(j)})}
\prod_{j=0}^{M} {\prod_{k=1}^{n_j} \left[\omega(\zeta,a_{k}^{(j)})\right]^{\beta_k^{(j)}}},
\label{eq:SC5}
\end{equation}
which combined with  (\ref{eq:lz}) gives
\begin{align}
\frac{dz}{d \zeta}
&=\mathcal{B} S(\zeta)
\prod_{j=0}^{M} \prod_{k=1}^{n_j} \left[\omega(\zeta,a_{k}^{(j)})\right]^{\beta_k^{(j)}},
\label{eq:SC6a}
\end{align}
where 
\begin{align}
S(\zeta)=\frac{  \omega_\zeta(\zeta,1)\omega(\zeta, -1)-\omega_\zeta(\zeta,-1)\omega(\zeta,1)}{\prod_{j=1}^{M}\omega(\zeta,\gamma_{1}^{(j)}) \omega(\zeta,\gamma_{2}^{(j)})}.
\label{eq:S1}
\end{align}
In (\ref{eq:SC6a}) a constant factor ($-\mathrm{i})$ has been absorbed into $\cal B$.

Upon integrating (\ref{eq:SC6a}), one then finds that the desired mapping $z(\zeta)$ is given by 
\begin{align}
z(\zeta)
&=\mathcal{A} +\mathcal{B} 
\int^\zeta S(\zeta') \prod_{j=0}^{M} \prod_{k=1}^{n_j} \left[\omega(\zeta',a_{k}^{(j)})\right]^{\beta_k^{(j)}} d\zeta'.
\label{eq:SC9}
\end{align}
where $\mathcal{A}$ and $\mathcal{B}$ are complex constants.
Recall  that the points $\{\gamma_{1}^{(j)}, \gamma_{2}^{(j)}\in C_j|j=1,...,M\}$  appearing in  the prefactor function $S(\zeta)$ in (\ref{eq:S1}) are  to be determined by  solving equation (\ref{eq:gj}), which in turn depends on the conformal moduli $\{\delta_j, q_j|j=1,...,M$\} of the domain $D_\zeta$. For a given polygonal domain $D_z$,  the conformal moduli of the domain $D_\zeta$ are not known  {\it a priori} and must be determined simultaneously with  the  other parameters appearing in (\ref{eq:SC6a}), namely, the prevertices $\{a_k^{(j)}\in C_j|j=0,1,...,M; k=1,...,n_j\}$ and the constant $\mathcal{B}$. Solving this \emph{parameter problem} \cite{Driscoll} in general is a very difficult task. 

Fortunately, in many applications, the specific details of the target polygonal domain (e.g., the areas and centroids of the polygons  $P_j$, and  the lengths of their respective edges)  need not be known {\it a priori}.
In such cases, 
one can freely specify the  domain $D_\zeta$, all prevertices on the unit circle, and all but two prevertices on each inner circle $C_j$, and then solve the reduced parameter problem associated with the orientation of the various polygons and  the univalence of the mapping function $z(\zeta)$. For instance,  
 formulae  (\ref{eq:S1}) and (\ref{eq:SC9}) were recently used by Green \& Vasconcelos \cite{GreenVasconcelos2013} to construct a conformal mapping from the circular domain $D_\zeta$ to a degenerate polygonal domain consisting of a horizontal strip with $M$ vertical slits in its interior (this conformal mapping corresponds to the complex potential for multiple steady bubbles in a Hele-Shaw channel).

\section{Representation formulae using other canonical slit domains}
\label{sec:6}

The procedure described in the previous section can be readily extended to other  rectilinear slit domains $D_\lambda$ in the subsidiary $\lambda$-plane, so long as the corresponding slit map $\lambda(\zeta)$ is known explicitly. For each choice of domain $D_\lambda$, a specific formula results for the prefactor  $S(\zeta)$ appearing in  conformal mapping (\ref{eq:SC9}). As an illustration of our procedure,
we derive  below the respective expressions for the function $S(\zeta)$ associated with two of the five canonical slit domains listed in the book of Nehari \cite{Nehari}, namely:  i)  a circular disk with $M$  concentric circular-arc slits; and ii) an unbounded radial slit domain obtained by excising from the entire plane $M+1$  rectilinear slits pointing toward the origin. (Other canonical rectilinear slit  domains  can be treated in similar manner.) 

Let us first  discuss the case of  an auxiliary slit domain consisting of a disk with concentric circular-arc slits, as originally considered by Crowdy  \cite{BoundedMCSC}.
Here
the  function 
\begin{align}
\eta(\zeta)=\frac{ \omega(\zeta,\alpha)}{|\alpha| \omega(\zeta,\bar\alpha^{-1})},
\label{eq:eta}
\end{align}
for $\alpha\in D_\zeta$,  maps the circular domain $D_\zeta$ onto 
the unit disc with $M$ concentric circular slits, where the point $\zeta=\alpha$ maps to the origin in the $\eta$-plane \cite{BoundedMCSC}. Thus, the logarithmic 
 transformation
 \begin{align}
 \lambda&=\log \eta(\zeta), 
  \label{eq:logeta}
 \end{align}
with an appropriate choice of branch cut from $\zeta=\alpha$ to $\zeta=1$,  maps $D_\zeta$ to a domain $D_\lambda$ in the $\lambda$-plane  consisting of a semi-infinite strip bounded from the right by the line ${\rm Re}[\lambda]=0$ and containing $M$ vertical slits in its interior, where the unit circle $C_0$ is mapped to the vertical edge of the strip and  the circles $C_j$, $j=1,...,M$, are mapped to  the vertical slits. As before, the points $\{\gamma_{1}^{(j)},\gamma_{2}^{(j)}\in C_j|j=1,...,M\}$  correspond to the preimages in the $\zeta$-plane of the end points of the slits in the $\lambda$-plane, so that $z_\lambda(\zeta)$ has simple poles at these points.  Since the point $\zeta=\alpha$ is a logarithmic singularity of the slit map $\lambda(\zeta)$, then $z_\lambda(\zeta)$ must have a simple zero at this point. 

Starting from (\ref{eq:SC4}) and in light of the preceding discussion, one readily concludes that in this case the correction function $R(\zeta)$  can be chosen as
\begin{align}
 R(\zeta)
 &= \frac{\omega(\zeta,\alpha)\omega(\zeta,\bar\alpha^{-1})}{\omega(\zeta,\gamma_1^{(0)})^2},
 \label{eq:R3}
 \end{align}
 which has constant argument on the circles $C_j$, as follows from (\ref{eq:argQ}) and from the fact that $\bar \gamma_1^{(0)}=1/\gamma_1^{(0)}$.
Inserting (\ref{eq:R3})  into (\ref{eq:SC4}) and setting $\gamma_2^{(0)}=\gamma_1^{(0)}$
(recall that  these points are arbitrary),  we obtain
\begin{equation}
\frac{dz}{d\lambda}=\mathcal{B}
\frac{ \omega(\zeta,\alpha)\omega(\zeta,\bar\alpha^{-1})}{\prod_{j=1}^{M} \omega(\zeta,\gamma_{1}^{(j)}) \omega(\zeta,\gamma_{2}^{(j)})}
\prod_{j=0}^{M} {\prod_{k=1}^{n_j} \left[\omega(\zeta,a_{k}^{(j)})\right]^{\beta_k^{(j)}}},
\label{eq:SC5b}
\end{equation}
Using (\ref{eq:logeta}) and (\ref{eq:SC5b}), one finds that the derivative $z_\zeta(\zeta)$ can be rewritten  as in (\ref{eq:SC6a}),
where the prefactor function $S(\zeta)$ now reads
\begin{equation}
S(\zeta)=
\frac{ \omega_\zeta(\zeta,\alpha)\omega(\zeta,\bar\alpha^{-1})-\omega_\zeta(\zeta,\bar\alpha^{-1})\omega(\zeta, \alpha)}{\prod_{j=1}^{M} \omega(\zeta,\gamma_{1}^{(j)}) \omega(\zeta,\gamma_{2}^{(j)})},
\label{eq:S3}
\end{equation}
which recovers the result obtained  by Crowdy  \cite{BoundedMCSC}.

As a further illustration of the generality of our approach, consider  now the case where  the domain $D_\lambda$ consists of the entire $\lambda$-plane with $M+1$ radial slits, denoted by $L_j$, $j=0,1,...M$. Recall that  the corresponding slit map $\lambda(\zeta)$ in this case is given by  (\ref{eq:lQ}) which has a simple pole  at $\zeta=\beta$; hence $z_\lambda(\zeta)$ must have a double  zero at  this point.
Note furthermore that  the points $\{\gamma_{1}^{(j)},\gamma_{2}^{(j)}\in C_j|j=0,1,...,M\}$  
  must correspond to the preimages of the end points of the respective slits $L_j$. This implies, in particular, that the prime functions containing the points $\gamma_{1}^{(0)}$ and $\gamma_{2}^{(0)}$ in the numerator of (\ref{eq:SC4})  must  be replaced with identical terms in the denominator, since $z_\lambda(\zeta)$ must now have simple poles at these two points, as well as at $\gamma_{1}^{(j)}$ and $\gamma_{2}^{(j)}$, $j=1,...,M$. This can be accomplished  with an appropriate choice of the function $R(\zeta)$, which must also produce the required double  zero at $\zeta=\beta$. Indeed, these requirements can be satisfied by choosing 
\begin{align}
 R(\zeta)
 &= \left[\frac{\omega(\zeta,\beta)\omega(\zeta,\bar\beta^{-1})}{\omega(\zeta,\gamma_0^{(1)})\omega(\zeta,\gamma_0^{(2)})}\right]^2.
 \label{eq:RP}
 \end{align}
After inserting (\ref{eq:RP}) into (\ref{eq:SC4}) and applying the chain rule, one finds
that the prefactor  $S(\zeta)$ for this case is given by
\begin{equation}
S(\zeta)=
\frac{ T(\zeta)}{\prod_{j=0}^{M} \omega(\zeta,\gamma_{1}^{(j)}) \omega(\zeta,\gamma_{2}^{(j)})},
\label{eq:S2}
\end{equation}
where
\begin{align}
T(\zeta)&= \omega_\zeta(\zeta,\alpha)\left[\omega(\zeta, \beta)\omega(\zeta,\bar\alpha^{-1})\omega(\zeta, \bar\beta^{-1})\right]
-\omega_\zeta(\zeta,\beta)\left[\omega(\zeta, \alpha)\omega(\zeta,\bar\alpha^{-1})\omega(\zeta, \bar\beta^{-1})\right]\cr &+\omega_\zeta(\zeta,\bar\alpha^{-1})\left[\omega(\zeta,\alpha)\omega(\zeta, \beta)\omega(\zeta, \bar\beta^{-1})\right]-\omega_\zeta(\zeta,\bar\beta^{-1})\left[\omega(\zeta, \alpha)\omega(\zeta, \beta)\omega(\zeta,\bar\alpha^{-1})\right].
 \end{align}
Similar expressions for the prefactor  function  $S(\zeta)$ pertaining to  other  canonical  slit domains can be readily obtained, but further details will not be presented here.

It is to be emphasized that, in contrast to formula (\ref{eq:S1})  for the upper half-plane with radial slits, the prefactor functions $S(\zeta)$  obtained for other canonical slit domains have arbitrary parameters, e.g., the point $\alpha$ in (\ref{eq:S3}) and the points $\alpha$ and $\beta$ in (\ref{eq:S2}), in the interior of the domain $D_\zeta$. To avoid this extra unnecessary complication, the formulation given in \S\ref{sec:5}  should be preferred in applications; see \S\ref{sec:8} for further discussion on this point.

\section{Conformal mappings to unbounded polygonal domains}
\label{sec:7}

In this section we consider, for completeness, the problem of conformal mappings from the bounded circular domain $D_\zeta$ to unbounded multiply connected  polygonal regions, using the upper half-$\lambda$-plane with $M$ radial slits as our auxiliary rectilinear slit domain. 

Let the target  domain $D_z$  in the $z$-plane be the unbounded region exterior to  $M+1$ nonoverlapping polygons  $P_j$, $j=0,1,...,M$. We shall adopt the same notation as in Sec.~\ref{sec:2} to designate the vertices of the polygonal boundaries and 
the corresponding turning angles. Notice, however, that now we have
\begin{align}
\sum_{k=1}^{n_j}\beta_k^{(j)}=2,\qquad j=0,1,...,M.
\label{eq:binf}
\end{align}
Here we wish to obtain a conformal mapping, $z(\zeta)$, from a bounded circular domain $D_\zeta$ to the unbounded polygonal region $D_z$, where each  circle $C_j$, $j=0,1,...,M$, is mapped to a polygonal boundary $P_j$ and the point $\zeta=\zeta_\infty$ is mapped to infinity. Employing a procedure similar to that used in \S\ref{sec:5} for bounded polygonal domains, analogous formula for the mapping of unbounded polygonal regions can be readily obtained.

The starting point for constructing the desired  mapping is the  equation  
\begin{equation}
\frac{dz}{d\lambda}=\mathcal{B}R(\zeta)
\frac{\prod_{j=0}^{M} \prod_{k=1}^{n_j} \left[\omega(\zeta,a_{k}^{(j)})\right]^{\beta_k^{(j)}}}{\prod_{j=0}^{M} \omega(\zeta,\gamma_{1}^{(j)}) \omega(\zeta,\gamma_{2}^{(j)})},
\label{eq:SC8}
\end{equation}
which is the counterpart of  expression (\ref{eq:SC4}) used for bounded polygonal domains. Notice that in contrast with  (\ref{eq:SC4}),  the prime functions containing the points $\gamma_{1}^{(0)}$ and $\gamma_{2}^{(0)}$   appear in the denominator of  (\ref{eq:SC8}) because now $\sum_{k=1}^{n_0}\beta_k^{(0)}=2$.
As before, the points $\{\gamma_1^{(j)},\gamma_2^{(j)}\in C_j|j=1,...,M\}$ are   identified with the preimages in the $\zeta$-plane of the end points of the $M$ slits in the $\lambda$-plane, whereas $\gamma_1^{(0)}$ and $\gamma_2^{(0)}$ are arbitrary points on $C_0$ at our disposal.

It is also clear from previous discussions that $z_\lambda(\zeta)$ must have a double pole at $\zeta=\zeta_\infty$ and a simple zero at $\zeta=-1$.
These  requirements can be enforced by   choosing
 \begin{align}
 R(\zeta)
 &= \left[\frac{\omega(\zeta,-1)\omega(\zeta,\gamma_1^{(0)})}{\omega(\zeta,\zeta_\infty)\omega(\zeta,1/\bar\zeta_\infty)}\right]^2.
 \end{align}
After inserting this into (\ref{eq:SC8}) and setting  $\gamma_1^{(0)}=\gamma_2^{(0)}$, one finds
 \begin{equation}
\frac{dz}{d\lambda}=\mathcal{B}
\frac{\omega(\zeta,-1)^2}{\left[\omega(\zeta,\zeta_\infty)\omega(\zeta,1/\bar\zeta_\infty)\right]^2\prod_{j=1}^{M} \omega(\zeta,\gamma_{1}^{(j)}) \omega(\zeta,\gamma_{2}^{(j)})}
\prod_{j=0}^{M} \prod_{k=1}^{n_j} \left[\omega(\zeta,a_{k}^{(j)})\right]^{\beta_k^{(j)}}.
\label{eq:SC8b}
\end{equation}
Upon using (\ref{eq:lz}), this becomes
\begin{equation}
\frac{dz}{d \zeta}=\mathcal{B} S_\infty(\zeta)
\prod_{j=0}^{M} \prod_{k=1}^{n_j} \left[\omega(\zeta,a_{k}^{(j)})\right]^{\beta_k^{(j)}},
\label{eq:SC8c}
\end{equation}
where
\begin{align}
S_\infty (\zeta) =\frac{S(\zeta)}{\left[\omega(\zeta,\zeta_\infty)\omega(\zeta,1/\bar\zeta_\infty)\right]^2},
\label{eq:Sinf}
\end{align}
with $S(\zeta)$ as  given in (\ref{eq:S1}). 

After integrating (\ref{eq:SC8c}), one  finds that the conformal mapping $z(\zeta)$ from $D_\zeta$ to an unbounded  multiply connected polygonal region is given by the same integral expression (\ref{eq:SC9}) obtained for the case of bounded polygonal domains, the only difference being that  the prefactor  is now given by the function $S_\infty(\zeta)$ shown in (\ref{eq:Sinf}). This property was first noticed by Crowdy \cite{UnboundedMCSC},  who obtained a conformal mapping from $D_\zeta$ to an unbounded polygonal region by implicitly considering an auxiliary rectilinear slit domain consisting of a semi-infinite strip with vertical slits. As shown above,  relation (\ref{eq:Sinf}) holds irrespective of the choice of the rectilinear slit domain used to construct the corresponding  mapping formulas for bounded and unbounded  multiply connected  polygonal domains.

\section{Discussion}
\label{sec:8}

A general framework has been presented for constructing conformal mappings from a bounded circular domain $D_\zeta$ to a multiply connected polygonal region $D_z$ (either bounded or unbounded). A key ingredient in our scheme is the introduction of a conformal mapping from $D_\zeta$ to  a  rectilinear slit domain $D_\lambda$ in a subsidiary $\lambda$-plane. This allows us to write an explicit formula for the derivative $z_\lambda(\zeta)$, and hence for $z_\zeta(\zeta)$, in terms of the Schottky-Klein prime function associated with the domain $D_\zeta$. After integration, the desired conformal mapping $z(\zeta)$ is then obtained as an indefinite integral whose integrand consists of a product of powers of the Schottky-Klein prime functions and a prefactor function $S(\zeta)$ that depends on the choice of the rectilinear slit domain $D_\lambda$. 

An explicit formula for $S(\zeta)$ was derived by first considering the case where the rectilinear slit domain $D_\lambda$ consists of the upper half-plane with radial slits. The generality of our approach was subsequently demonstrated by obtaining alternative formulae for the prefactor function  $S(\zeta)$ pertaining to  other canonical  slit domains.
For a given polygonal domain $D_z$, these various formulas (once their associated  parameters have been determined) provide  different representations of the same conformal mapping $z(\zeta)$.

It is to be noted, however, that the formula for $S(\zeta)$ obtained by considering the upper half-plane with radial slits is arguably the simplest one, in the sense that the only unknown parameters are the zeros of the slit map $\lambda(\zeta)$, which can be numerically computed once the domain $D_\zeta$ is specified. 
By contrast, the corresponding formulae for $S(\zeta)$
obtained for other canonical slit domains have, in addition, one or more arbitrary parameters inside the domain $D_\zeta$. Although 
the function $S(\zeta)$ does not ultimately depend on the values of these parameters (except for an overall factor independent of $\zeta$; see Crowdy \cite{DarrenRHpaper}), the existence of arbitrary parameters  inside $D_\zeta$ may present an additional (and unnecessary) source of complication. This is particularly true in the case that
 a given target polygonal domain $D_z$ is specified, for here
 the conformal moduli of the domain $D_\zeta$ are not known {\it a priori} and hence the arbitrary parameters cannot be fixed beforehand.
  
 In light of the foregoing discussion, it can be argued that the mapping formula derived in \S\ref{sec:5} using the upper half-plane with radial slits should  be viewed as the natural extension to multiply connected polygonal domains of the standard Schwarz-Christoffel mapping from the upper half-plane to a simply connected polygonal region. It should   also be preferable in applications because of its simplicity.
 In this context, it is  worth noting that this mapping formula 
 was recently employed by Green \& Vasconcelos \cite{GreenVasconcelos2013} to construct exact solutions for multiple bubbles steadily moving in a Hele-Shaw channel. 
 It is  thus hoped that other problems involving multiply connected polygonal domains may  be conveniently tackled with the  formalism presented here.

\section*{Acknowledgments}
{The author wishes to thank C. C. Green and D. G. Crowdy for helpful discussions.  He is  also appreciative of  the hospitality of the Department of Mathematics at Imperial College London (ICL), where this research was carried out. Financial support from a scholarship from the Conselho Nacional de Desenvolvimento Cientifico e Tecnologico (Brazil) for a sabbatical stay at ICL is acknowledged.}



\begin{thebibliography}{}

\bibitem{Driscoll} Driscoll T, Trefethen LN. 2002 {\it Schwarz-Christoffel mapping Cambridge mathematical monographs}. Cambridge: Cambridge University Press.

\bibitem{DeLillo2004} DeLillo TK, Elcrat AR, Pfaltzgraff JA. 2004 Schwarz-Christoffel mapping of multiply-connected domains. {\it J. d'Analyse Math.} {\bf 94}, 17-47. (doi:
10.1007/BF02789040)

\bibitem{DeLillo2006} DeLillo TK. 2006 Schwarz-Christoffel mapping of bounded, multiply connected domains. {\it Comput. Methods Funct. Theory} \textbf{6}, 275-300. (doi:
10.1007/BF03321615)


\bibitem{BoundedMCSC} Crowdy DG. 2005 The {S}chwarz-{C}hristoffel mapping to bounded multiply connected polygonal domains. {\it Proc. Roy. Soc. A} {\bf 461}, 2653-2678. (doi: 10.1098/rspa.2005.1480)

\bibitem{UnboundedMCSC} Crowdy DG. 2007 Schwarz-{C}hristoffel mappings to unbounded multiply connected polygonal regions. {\it Math. Proc. Cambridge Philos. Soc.} {\bf 142}, 319-339. (doi: 10.1017/S0305004106009832)


\bibitem{Fokas} Ablowitz MJ, Fokas AS. 1997 {\it Complex Variables: Introduction and Applications}. Cambridge: Cambridge University Press.

\bibitem{Baker} Baker HF. 1897 {\it Abelian functions: Abel's theorem and the allied theory of theta functions}. Cambridge: Cambridge University Press.

\bibitem{KRtheory}  Crowdy DG, Marshall JS. 2005 Analytical formulae for the Kirchhoff-Routh path function in multiply connected domains. {\it Proc. R. Soc. A} \textbf{461}, 2477-2501. (doi:10.1098/rspa.2005.1492)

\bibitem{Hejhal} Hejhal DA. 1972 {\it Theta functions, kernel functions and {A}belian integrals}. Memoirs of the American Mathematical Society, Vol. 129, American Mathematical Society, Providence. 

\bibitem{CM2006}
Crowdy DG, Marshall JS. 2006 Conformal mappings between canonical multiply connected domains. {\em Comput. Methods Funct. Theory} \textbf{6}, 59-76.

\bibitem{DeLillo2010} DeLillo TK, Kropf HK. 2010 Slit maps and Schwarz-Christoffel maps for multiply connected domains. {\it Electron. Trans. Numer. Anal.} {\bf 36}, 195-223.

\bibitem{GreenVasconcelos2013} Green CC, Vasconcelos GL. 2014 Multiple steadily translating bubbles in a Hele-Shaw channel. {\it Proc. R. Soc. A} {\bf 470}, 20130698. (doi:10.1098/rspa.2013.0698)

\bibitem{Nehari} Nehari Z. 1952 {\it Conformal Mapping}, McGraw-Hill, New York.


\bibitem{DarrenRHpaper} Crowdy DG. 2009 Explicit solution of a class of {R}iemann-{H}ilbert problems. {\it Ann. Univ. Paedagog. Crac. Stud. Math.} {\bf 8}, 5-18.


\end{thebibliography}
\end{document}